\documentclass{iopjournal}

\begin{document}
\articletype{Paper} 

\title{Depth and slip ratio dependencies of friction for a sphere rolling on a granular slope}

\author{Takeshi Fukumoto$^{1,*}$\orcid{0009-0003-9234-1274}, Hiroyuki Ebata$^1$\orcid{0000-0002-5370-631X}, Ishan Sharma$^2$\orcid{0000-0002-8447-4857} and Hiroaki Katsuragi$^{1,3}$\orcid{0000-0002-4949-9389}}

\affil{$^1$Department of Earth and Space Science, The University of Osaka, 1-1 Machikaneyama Toyonaka, Osaka, Japan}

\affil{$^2$Department of Mechanical Engineering, and Space, Planetary \& Astronomical Sciences and Engineering, Indian Institute of Technology Kanpur, India}

\affil{$^3$Department of Space, Planetary {\&} Astronomical Sciences and Engineering, Indian Institute of Technology Kanpur, India}

\email{t.fukumoto@ess.sci.osaka-u.ac.jp}

\keywords{granular, friction, rolling}

\begin{abstract}
We experimentally investigate the dynamics of a sphere rolling down a granular slope by varying the initial velocity, slope angle, and sphere density. The results show that the sphere rolls down with constant deceleration while sinking into the granular bed. $\delta/R$ (the sinking depth $\delta$ normalized to the sphere radius $R$) is scaled by the sphere density normalized by the bulk density of the granular layer. To evaluate the translational energy dissipation, we introduce an effective friction coefficient $\mu_\mathrm{d}$. We demonstrate that $\mu_\mathrm{d}$ decreases with increasing the slope angle and the slip ratio. Furthermore, systematic measurements over a wide range of sphere densities reveal that $\mu_\mathrm{d}$ increases linearly with $\delta/R$ : $\mu_\mathrm{d}=\beta(\delta/R)+\mu_0$. The value of $\mu_0$ is linearly decreasing with slip ratio and its coefficient $\beta(\simeq0.41)$ does not vary significantly. The results suggest that the normalized depth and slip ratio determine the effective friction of a rolling sphere.

\end{abstract}

\section{Introduction}

Rolling motion on granular surfaces is observed in many contexts, including planetary science, biology, and engineering. 
In planetary settings, examples include boulder falls on the lunar surface~\cite{Bickel_2021, Ikeda_2022} and rockfalls on Earth~\cite{Cui_2017}. In biology, dung beetles~\cite{dungbeetle} and pillbugs roll objects forward~\cite{pill_bug}. In engineering, the rolling behavior of tires on granular surfaces is a significant practical issue. The relevant examples include truck escape ramps~\cite{escape_2019} and planetary rovers. For instance, the Mars rover, Spirit eventually became stuck due to a lack of traction caused by the stirring of sand during planetary exploration~\cite{Sanderson_2010}. Thus, it is essential to elucidate the interaction between the deforming ground and the rolling object that is common to these phenomena.

It is a simple setup in which an object rolls across deformable granular ground, yet surprisingly little scientific knowledge is available on this subject. Many works reported the friction and resistance between deformable granular media and solid objects. However, most research has focused on resistance and friction when moving in a translational direction without rotation.

For the case of pure translational motion relating to drag force and the resistance of granular bed, many investigations were reported. 
For instance, the drag force during vertical impact cratering into granular media was analyzed~\cite{Katsuragi_2007, Okubo_2022}. 
Penetration drag at constant velocity was also examined~\cite{Seguin_2013, Brzinski_2013}. It was shown that, depending on the object’s shape, the resistance was proportional to either the square or the cube of its radius~\cite{Brzinski_2013}. Iikawa and Katsuragi explored the effects of particle adhesion (wet granular), inter-particle friction, and shape of intrusion on penetration resistance~\cite{iikawa}. 
Objects moving at constant velocity through granular media were also examined~\cite{Ding_2011, Potiguar_2013, Seguin_2019}, where it was found that horizontally moving objects experienced both horizontal resistance and vertical lift forces. 
Motions with acceleration~\cite{Gondret_2022} or deceleration~\cite{Pacheco_2009} in granular media were also considered. 
Translational motion accompanied by a certain degree of sinking into the granular surface, either at an imposed constant velocity~\cite{Liefferink_2018, Rinse_2020, Carvalho_2024} or under gravity along granular slopes~\cite{Crassous_2017}, was likewise reported. 
These studies demonstrated that deformation of the granular medium affected the friction between the object and the surface~\cite{Liefferink_2018, Rinse_2020, Crassous_2017}.

In the case of combination of rotating and translating motion, some studies were conducted. 
Penetration with rotation at constant velocity was also investigated~\cite{Jung_2017}, where it was found that rotational motion reduces the drag force. Impact cratering with rolling was likewise examined~\cite{Jung_2017}, and the results showed that rotation increases the sinking depth. Motion with rolling at various slip ratios was explored~\cite{Seguin_2022}, demonstrating that rotational motion reduced the drag force up to a certain rotational velocity. 
The characteristic of these studies~\cite{Darbois_2017, Jung_2017, Seguin_2022} is that they are externally driven and objects move at a constant speed. Some studies addressed the behavior of a sphere rolling on granular surface without its translational or rotational speed being controlled by a motor. In this perspective, cases such as rolling on a flat horizontal granular surface~\cite{Stefaan_2017}, rolling down a granular slope with deceleration~\cite{Blasio_2009}, rolling down with acceleration~\cite{Texier_2018}, and rolling up a granular slope~\cite{Fukumoto_2024} are of direct relevance to the present study.

Wal et al.\ examined the friction between a flat horizontal granular surface and a rolling basketball or medicine ball using both numerical and experimental approaches~\cite{Stefaan_2017}. Although they suggested that the friction coefficient depends on the sinking depth, no quantitative relationship was obtained. Fukumoto et al.\ analyzed the dynamics of a sphere rolling up a granular slope~\cite{Fukumoto_2024} and found an empirical quantitative relation in which the friction coefficient $\mu_\mathrm{d}$ is proportional to the sinking depth normalized to the radius of the sphere $\delta/R$, $\mu_\mathrm{d}=0.49(\delta/R)$. However, the range of applicability of this empirical relation, namely its dependence on the slope angle (including both rolling up and down), was not clarified. In fact, the friction coefficient observed when a sphere rolled down a granular slope~\cite{Blasio_2009}, $\mu_\mathrm{d}\simeq0.45$ was larger than the value expected from this empirical law, $\mu_\mathrm{d}=0.49(\delta/R)$, because the normalized sinking depth was $\delta/R\simeq0.3$. Besides, Blasio and Saeter~\cite{Blasio_2009} conducted their experiments only within a narrow range of sphere densities (they only use glass and steel) and slope angles (14.5$^\circ$–20.5$^\circ$). Thus, the dependencies on density and slope angle were not fully clarified.

In this study, we experimentally investigate the effect of a wide range of sphere densities and slope angles on the decelerative dynamics and friction of a sphere rolling down a granular slope. In other words, applicable range of the previously reported empirical law is experimentally evaluated. The results suggest that the normalized depth and slip ratio determine the effective friction of a rolling sphere. The friction is characterized from kinematic data obtained in a dedicated experimental setup. The details of the experimental apparatus, parameters, and conditions are described in Sec.~\ref{sec_Experiment}. The experimental results analyzed using the analytical method outlined in Sec.~\ref{sec_Analysis} are presented in Sec.~\ref{sec_result}. A discussion introducing the frictional framework is given in Sec.~\ref{sec_Discussion}, and the Conclusion is provided in Sec.~\ref{sec_conclusion}

\section{Experiment}\label{sec_Experiment}
\begin{figure}
    \centering
    \includegraphics[width=\columnwidth]{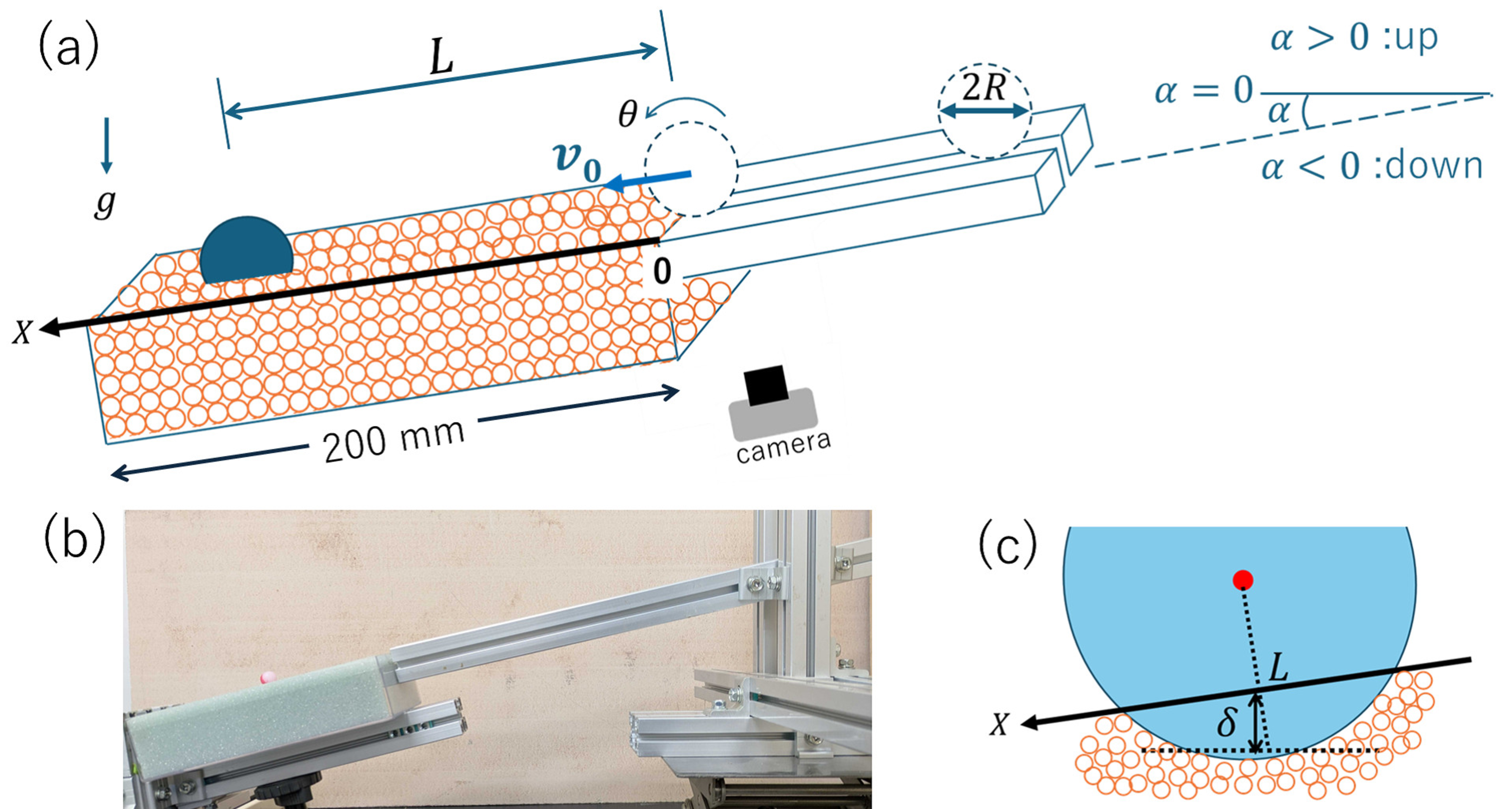}
        \caption{(a) Schematic view of the experimental setup. By rolling down the rail made of aluminium, the sphere of the radius $R$ obtains initial translational velocity $v_0$. The rolling posture $\theta$ is measured counterclockwise. The sphere rolls down the granular slope of the angle $\alpha$. $\alpha<0$ and $\alpha>0$ (obtained by Fukumoto et al.~\cite{Fukumoto_2024}) are defined as downhill and uphill, respectively. (b) The view of the experimental setup. (c) The sphere moves in the $X$ direction by a distance $L$ and sinks by a depth $\delta$ when it stops. [(a), (c)] The size ratio between the sphere and the glass beads is not to scale.}
    \label{fig:experiment}
\end{figure}

To obtain the dynamics of the sphere rolling down a granular slope, we build the experimental setup as shown in Fig.~\ref{fig:experiment}~(a) and (b). The sphere is released at various heights on the aluminium rail (rail width: 6~$\mathrm{mm}$) to obtain the initial velocity $v_0$ (0.32–0.95~$\mathrm{m/s}$ as shown in tab.~\ref{tab:v0}) at $X=0$, and then enters the granular slope. The initial velocity $v_0$ is varied by changing the initial release position of the sphere. We intentionally adjust $v_0$ so that the sphere would come to rest within the glass-bead box mentioned later in this section. In addition, the minimum initial velocity is limited such that the sphere rolls over a sufficient distance ($>$ 20~mm) to minimize wall effects. Because the deceleration depends on the density of the sphere, it is necessary to adjust the initial velocity $v_0$ to some extent. The $X$ axis is taken along the granular slope. The spheres used in this experiment are made of polyethylene, polyacetal, glass, and alumina ceramic (mainly supplied by As-one). Their densities are $\rho_\mathrm{s} = 930$, 1400, 2600, and 3900~$\mathrm{kg/m^3}$, respectively. All spheres are sufficiently rigid to neglect deformation and have the identical surface static friction coefficients against the fixed glass beads surface, 0.12. This value is measured by the critical angle on which the sphere start to roll down on the fixed glass beads surface.
The spheres used in this study are the same as those used in the previous uphill study~\cite{Fukumoto_2024}. However, the upper density limit is set to that of alumina ceramic since it was impossible to perform image analysis for a stainless steel sphere due to the large splash associated with its deep sinking~\cite{Fukumoto_2024}. The lower density limit corresponds to that of polyethylene sphere because the lighter density (e.g. polystyrene) sphere cannnot be ignored in the air resistance. The density range is determined based on these criteria. All these spheres have an identical radius of $R = 6.35$~mm. The spheres used in the experiments have diameters approximately ten times larger than the particle size, ensuring a scale much larger than that of individual grains. The box (width: 200~mm, length: 100~mm, height: 50~mm) filled with glass beads is inclined by an angle $\alpha$. The $\alpha$ is varied as $\alpha \simeq -5^{\circ}$, $-10^{\circ}$, $-15^{\circ}$, or $-20^{\circ}$. The typical diameter of the glass beads used in this experiment is 0.8~mm (770–910~{\textmu}m, AS-One, BZ-08). The employed diameter should be large enough to neglect air drag and cohesive forces. At the same time, the size ratio between the rolling sphere and the glass beads should be large. Considering this compromise, the sizes of the materials used were determined. The true density of the glass beads is $\rho = 2500~\mathrm{kg/m^3}$. The bulk density of the granular layer in this experiment is estimated as $\rho_\mathrm{g} = 1600 \pm 40~\mathrm{kg/m^3}$, which corresponds to the packing fraction $\phi$ given by $\rho_\mathrm{g} = \phi \rho$, resulting in $\phi = 0.64 \pm 0.02$. Each experimental condition has five experimental runs to evaluate the reproducibility. Before each run, the granular layer is completely removed from the box and the box is refilled with glass beads. To capture the dynamics of the sphere, we use a high-speed camera (Omron Sentech, STCMCCM401U3V) operating at 200 fps with a spatial resolution of 0.13~mm/pixel (image size: 2048 × 700 pixels). From the acquired images, we measure the instantaneous position and rolling posture of the sphere, the sinking depth $\delta$, the maximum travel distance $L$, and the slip ratio $s$ ($s=1-L/R\theta_\mathrm{stop}$, where $\theta_\mathrm{stop}$ is the total rotational angle) as shown in Fig.~\ref{fig:experiment}~(c).

\begin{table}
\caption{$v_0$ [m/s] at each $\alpha$ and the sphere.}
\centering
\begin{tabular}{l c c c c}
\hline
 & polyethylene & polyacetal & glass & alumina ceramic \\
\hline
$\alpha = -5^{\circ}$ (-0.087 rad) & 0.32-0.55 & 0.32-0.42 & 0.35-0.40 & 0.35-0.45\\ 
    $\alpha = -10^{\circ}$ (-0.17 rad) & 0.45-0.7 & 0.45-0.78 & 0.56-0.9 & 0.6-0.95 \\ 
    $\alpha = -15^{\circ}$  (-0.26 rad) & 0.37-0.60 & 0.36-0.67 & 0.46-0.7 & 0.6-0.85 \\ 
    $\alpha = -20^{\circ}$  (-0.35 rad) & 0.35-0.5 & 0.43-0.6 & 0.44-0.7 & 0.5-0.77 \\ 
\hline
\end{tabular}
\label{tab:v0}
\end{table}

\section{Analysis}\label{sec_Analysis}
\begin{figure}
    \centering
    \includegraphics[width=.8\columnwidth]{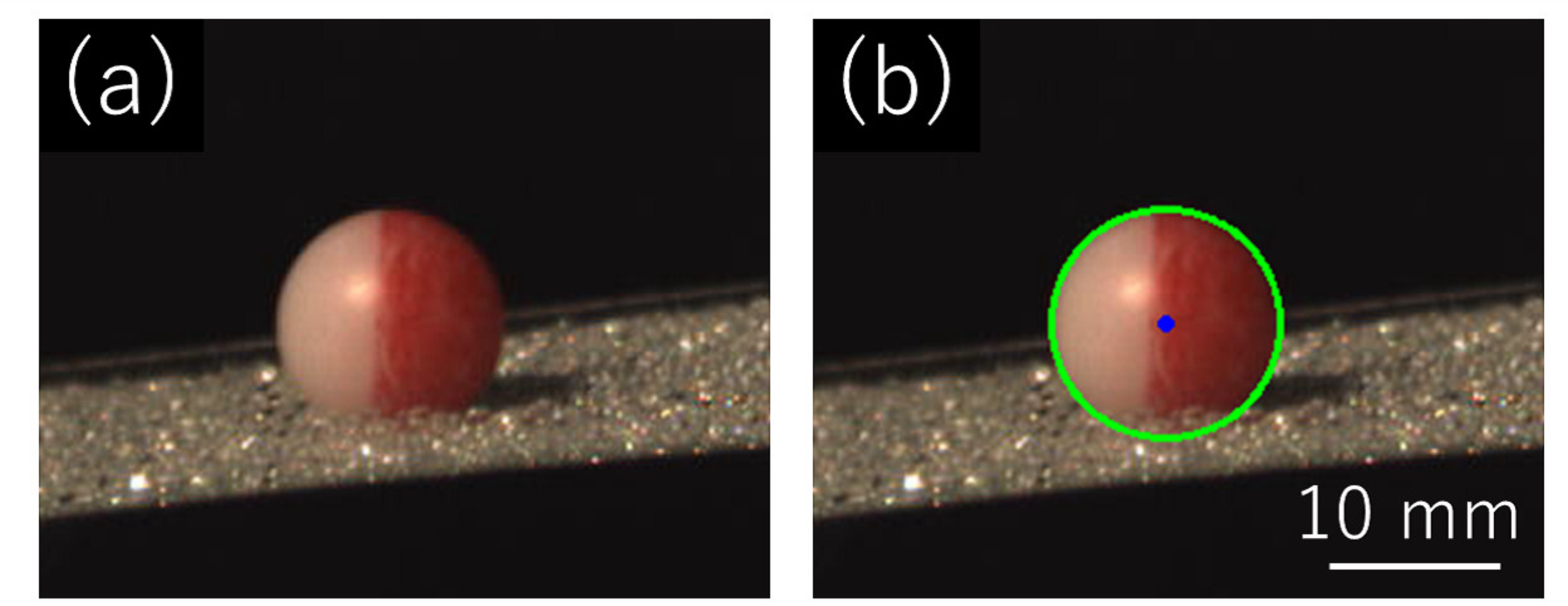}
    \caption{(a) The raw image of polyethylene sphere rolling down a slope of $\alpha = -5^{\circ}$. The hemispherical part of the sphere is
colored in red to measure the rolling posture. (b) The image after image analysis of (a). The green curve and the blue point indicate the detected circle and the center of the sphere respectively.}
    \label{fig:image}
\end{figure}

The typical examples of actual movies are provided as supplementary material. As shown in Fig.~\ref{fig:image}~(a), a hemisphere of each sphere, except for the glass sphere, is painted red to detect the rolling orientation. 
Due to the surface condition of the glass sphere, it is not possible to paint it red with a permanent marker. The total rotational angle during $0\leq x \leq L$ is defined as $\theta_\mathrm{stop}$ (including  the number of full rotations (each corresponding to $2\pi$)) and is measured in the counterclockwise direction by detecting the slope of the boundary between the white and red regions except for the glass sphere. To simplify the analysis, we evaluate only the initial and final images, together with the total number of rotations, which are manually counted. Only from the initial and final angles and rotation number, we calculate the total rotational angle $\theta_\mathrm{stop}$. Note that we do not measure the time-resolved rotational angle. By detecting the center of the sphere in each image, as shown in Fig.~\ref{fig:image}~(b), we obtain the sinking depth $\delta$ in the direction of gravity and the translational trajectory of the sphere, $X(t)$, where $t$ is the elapsed time after entering the granular slope. 

\section{Result}\label{sec_result}
\subsection{depth}\label{subsec_depth}

\begin{figure}
    \centering
    \includegraphics[width=\columnwidth]{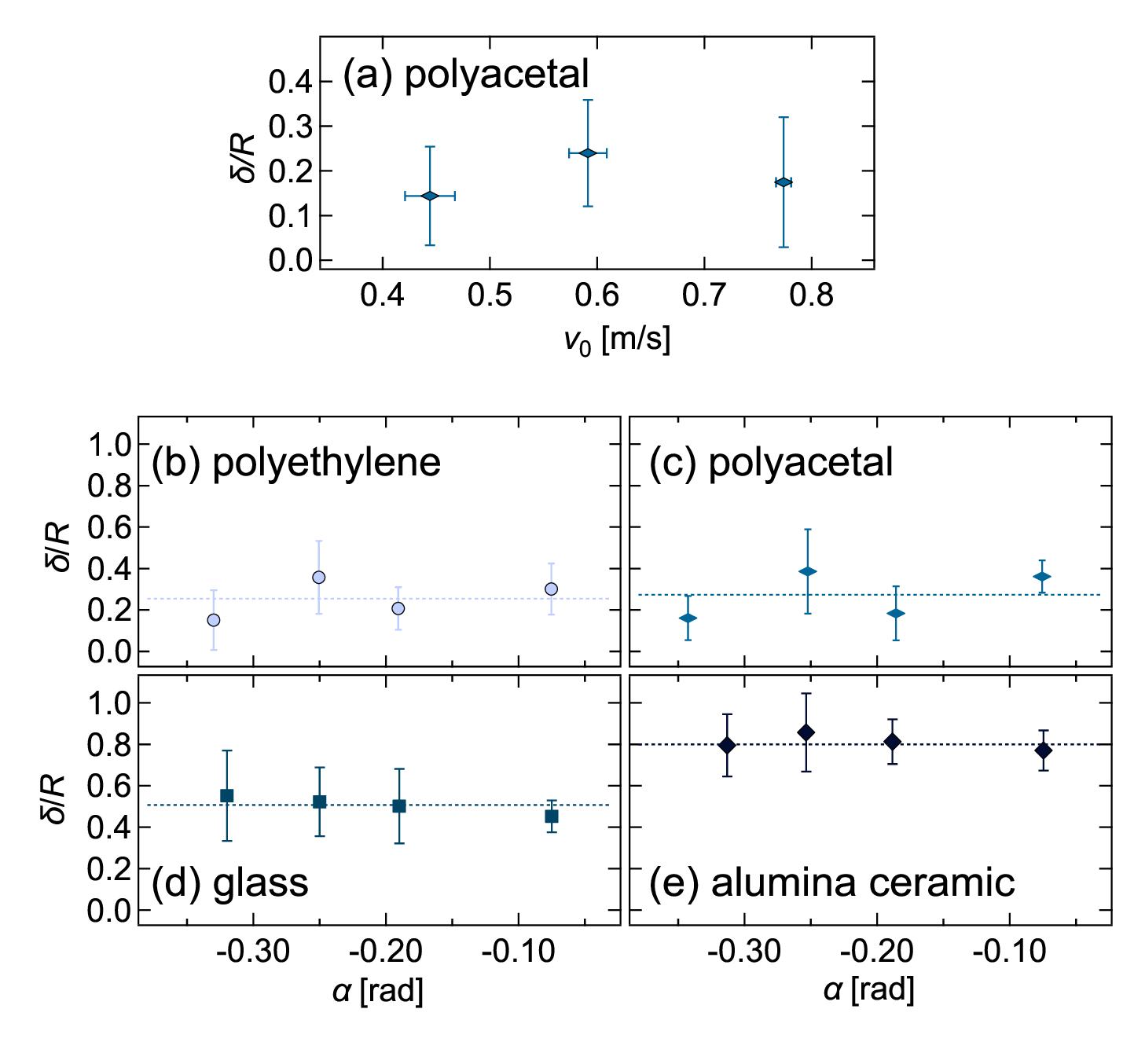}
    \caption{(a) The relation between depth normalized by radius $\delta/R$ and initial velocity $v_0$ for polyacetal at $\alpha\simeq-10^{\circ}$ taken as a representative data. Error bars indicate the standard deviation of 5 experimental runs. [(b)-(e)]~The relation between $\delta/R$ and angle $\alpha$ [(b) polyethylene (c) polyacetal (d) glass (e) alumina ceramic]. Error bars indicate the standard deviation of various initial velocity $v_0$ and repeated experimental runs. The dashed lines indicate the average of all data in each sphere.}
    \label{fig:dvsa}
\end{figure}

\begin{figure}
    \centering
    \includegraphics[width=\columnwidth]{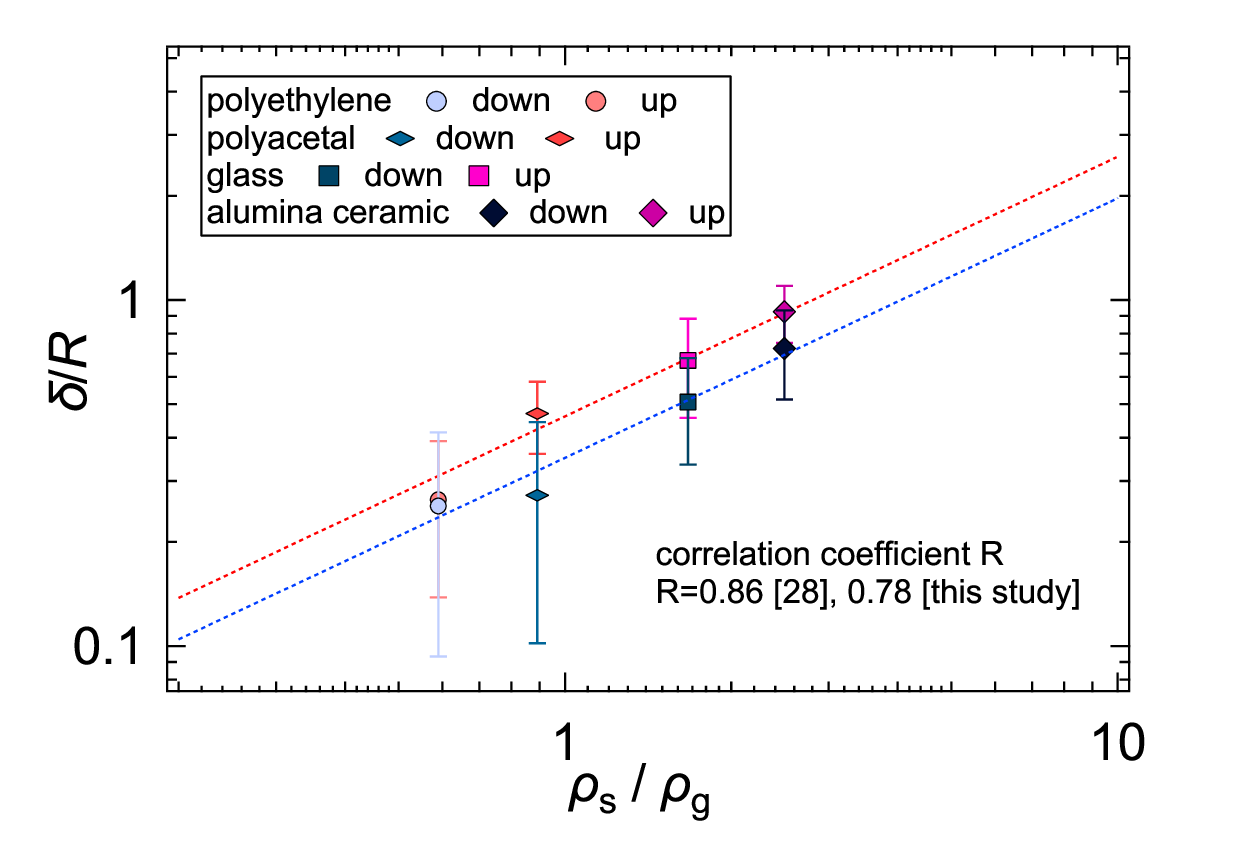}
  \caption{The double logarithmic plot of $\delta/R$ vs $\rho_\mathrm{s}/\rho_\mathrm{g}$ for both downhill and uphill. We use the data (uphill) obtained by Fukumoto et al.~\cite{Fukumoto_2024}. Error bars indicate the standard deviation of various $\alpha, v_0$ conditions. The blue and red dashed line indicate the scaling relation [Eq.~(\ref{eq:depth_fitting})] with $C_\mathrm{\rho}=0.38$ (R=0.78) and 0.46 (R=0.86), respectively. Here, R is the correlation coefficient.}
    \label{fig:fitting_depth}
\end{figure}

The measured sinking depth normalized by the sphere radius, $\delta/R$, is plotted in Fig.~\ref{fig:dvsa}. As shown in Fig.~\ref{fig:dvsa}~(a), $\delta/R$ is almost independent of $v_0$. There is an increasing trend beyond the error bars at $\alpha=-15^\circ$, but otherwise it is almost constant within the error bars regardless of initial velocity $v_0$.
Therefore, in Fig.~\ref{fig:dvsa}~(b)–(e), the average is taken over data with different $v_0$, and it can be seen that it does not depend on $\alpha$, similar to $v_0$. As an overall trend, the average value of $\delta/R$ depends on the density of the sphere, $\rho_\mathrm{s}$. As shown in Fig.~\ref{fig:fitting_depth}, $\delta/R$ and $\rho_\mathrm{s}/\rho_\mathrm{g}$ exhibit the positive correlation. The density dependence of sinking depth has been investigated previously~\cite{Uehara_2003,Texier_2018,Fukumoto_2024}. For a sphere dropped onto a granular surface without initial velocity, the normalized sinking depth was found to scale with the density ratio $(\rho_\mathrm{s}/\rho_\mathrm{g})$ as obtained by Uehara et al.~\cite{Uehara_2003}, 
\begin{equation}
\frac{\delta}{R}= C_{\rho} \left( \frac{\rho_\mathrm{s}}{\rho_\mathrm{g}} \right)^{3/4},
\label{eq:depth_fitting}    
\end{equation}
where $C_{\rho}$ is a dimensionless constant that depends on the friction coefficient between grains. Uehara et al.\ reported $C_{\rho}=0.51$ for a flat horizontal glass-bead surface~\cite{Uehara_2003}, while Texier et al.\ obtained $C_{\rho}=0.61$ for a granular slope~\cite{Texier_2018}. Although the value of $C_{\rho}$ differs, this scaling law [Eq.~(\ref{eq:depth_fitting})] can also be applied to the sinking depth after a sphere rolled up a granular slope ($C_{\rho}=0.46\pm0.05$)~\cite{Fukumoto_2024}. When 
assuming that this scaling relation [Eq.~(\ref{eq:depth_fitting})] is also applicable to the present study, we find that the relation is satisfied with $C_{\rho}=0.38\pm0.06$, as shown in Fig.~\ref{fig:fitting_depth}. Therefore, this scaling law is valid for spheres rolling on granular surfaces, and also the coefficient $C_{\rho}$ does not vary significantly between the rolling-down and rolling-up cases~\cite{Fukumoto_2024}.

\subsection{translational motion} \label{sec_translation}
\begin{figure}
    \centering
    \includegraphics[width=\columnwidth]{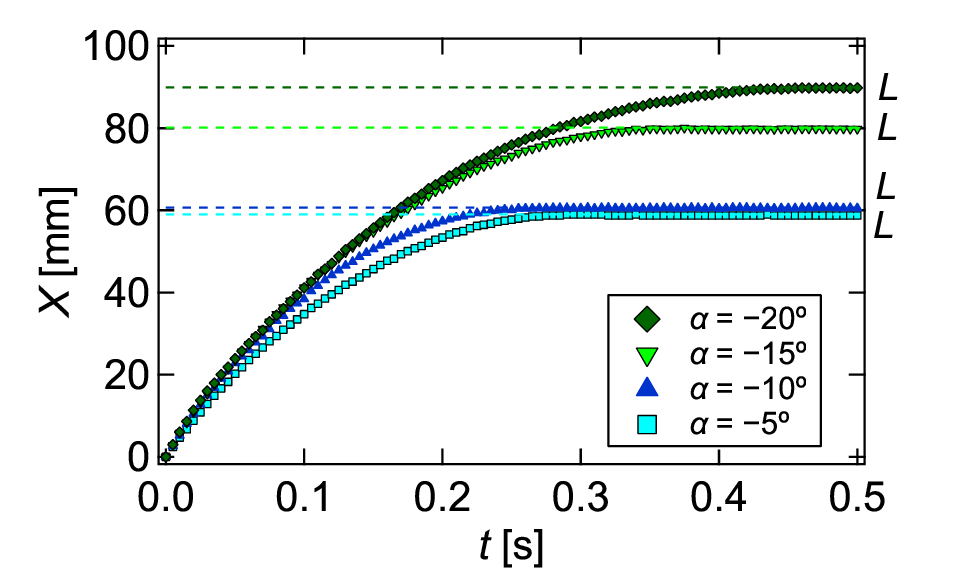}
    \caption{The translational $X$ position as a function of time $t$ for the polyacetal sphere [$v_0 \simeq 0.45~\mathrm{m/s}$] taken as a representative data. $L$ indicates the maximum travel distance at each $\alpha$.}
    \label{fig:step}
\end{figure}

The measured instantaneous position of the sphere, $X(t)$, for the polyacetal sphere is shown in Fig.~\ref{fig:step}, as a representative case. As seen in Fig.~\ref{fig:step}, $L$, defined as the maximum of $X(t)$, decreases with $\alpha$. 
These tendencies are also observed for the other spheres.  

\begin{figure}
    \centering
    \includegraphics[width=\columnwidth]{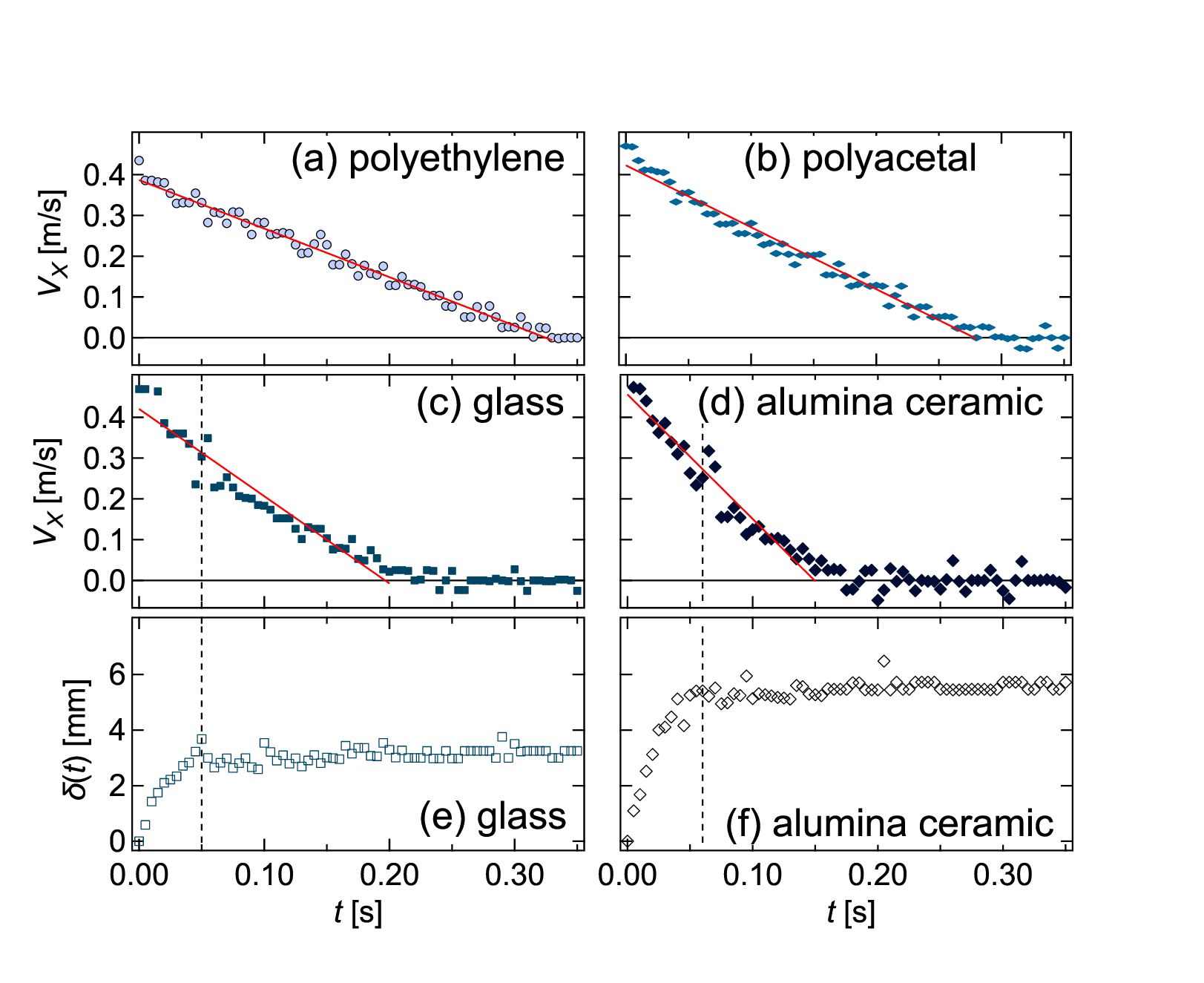}
    \caption{[(a)-(d)] The translational velocity $V_X$ as a function of time $t$ at $\alpha\simeq-5^{\circ}$. The red lines indicate the least-square line fitting from $t=0$ to the time when $V_X = 0$ [(a) polyethylene (b) polyacetal (c) glass (d) alumina ceramic]. [(e), (f)] The sinking depth $\delta(t)$ as a function of $t$. The dashed line indicates the time when $\delta$ becomes constant [(e) glass (f) alumina ceramic].}
    \label{fig:vt}
\end{figure}

We then analyze the velocity $V_X(t) =dX(t)/dt$ for the case of $\alpha\simeq-5^\circ$, as shown in Fig.~\ref{fig:vt}. $V_X(t)$ decreases linearly with $t$. Note that compared to (a) and (b) in Fig.~\ref{fig:vt}, the linear tendency is weaker for spheres with higher densities such as (c) and (d) in Fig.~\ref{fig:vt}. This is likely because the sinking depth is greater and it takes longer to reach the certain constant sinking depth $\delta$ as shown in Fig.~\ref{fig:vt}~(e, f). After $\delta$ becomes constant, we consider the motion to be characterized by constant deceleration during the rolling because the trend in late stage is linear as shown in Fig.~\ref{fig:vt}~(c, d). This linear trend is consistently observed for other values of $\alpha$. The slope of the fitting line becomes steeper as the sphere density increases. 
The tendency of constant acceleration (or deceleration) has also been reported in previous studies~\cite{Stefaan_2017,Blasio_2009,Texier_2018,Fukumoto_2024}, and appears to be a general characteristic of spheres rolling on deformable granular surfaces. We discuss this translational motion in terms of friction in Sec.~\ref{sec_Discussion}.

\subsection{rotational motion} \label{sec_rotation}
\begin{figure}
    \centering
    \includegraphics[width=\columnwidth]{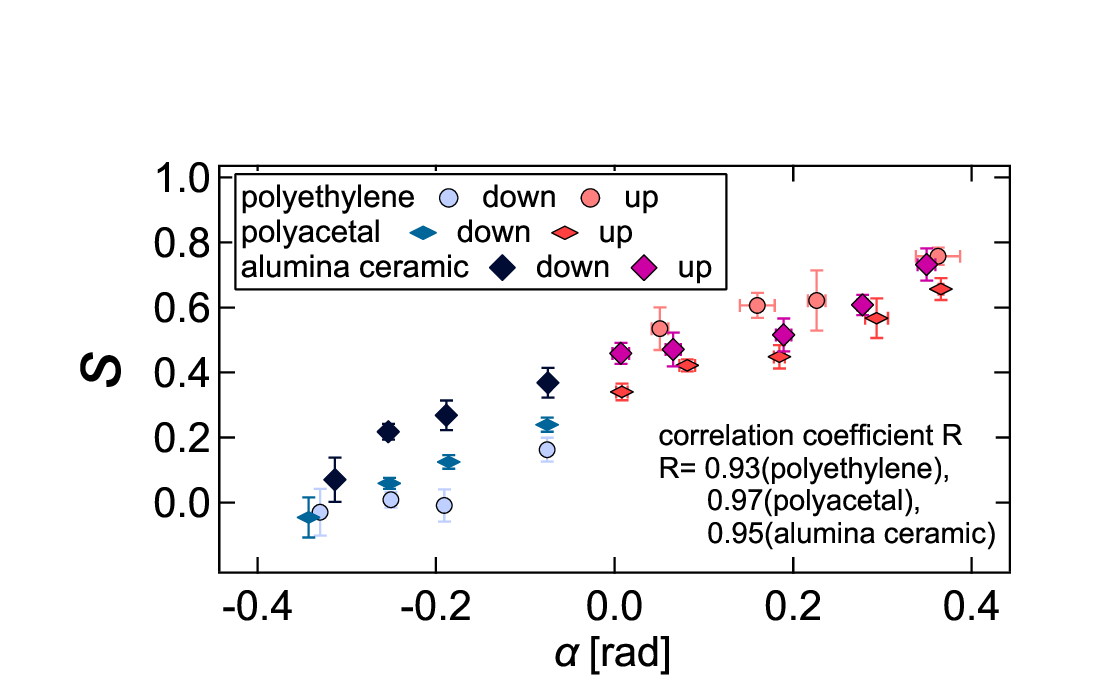}
    \caption{The relation between slip ratio $s$ and angle $\alpha$. $\alpha<0$ and $\alpha>0$ correspond to rolling down and up cases respectively. We use the data ($\alpha>0$) obtained by Fukumoto et al.~\cite{Fukumoto_2024}. Error bars indicate the standard deviation of all the data. R=0.93(polyethylene), 0.97(polyacetal), 0.95(alumina ceramic), R: the correlation coefficient.}
    \label{fig:rotation}
\end{figure}

To examine the relationship between the maximum travel distance $L$ and the total rotation angle $\theta_\mathrm{stop}$ before stopping, we introduce the slip ratio $s$, defined as
\begin{equation}
    (1-s)R\theta_\mathrm{stop} = L.
    \label{eq:slip}
\end{equation}
The slip ratio $s$ is computed by substituting the measured values of $L$ and $\theta_\mathrm{stop}$ into Eq.~(\ref{eq:slip}). Note that when the sphere is rolling down a slope, the translational and rotational motions stop almost simultaneously. As shown in Fig.~\ref{fig:rotation}, $s$ increases with $\alpha$, where the definition of $\pm\alpha$ corresponds to rolling up $(+)$ and rolling down $(-)$. (For rolling up $(+)$, we use the data obtained by Fukumoto et al.~\cite{Fukumoto_2024}.) The $s$ and $\alpha$ show the positive correlation as an overall trend. Moreover, the variation of $s$ appears to be roughly continuous for each sphere as $\alpha$ changes. The mechanism governing slipping behavior is complex. The shear imposed by the sphere rolling develops a localized shear zone called shear banding structure. This related phenomenon is reported by~\cite{shear} The relation between shear banding structure and slipping behavior should be revealed to understand the fundamental physical process of energy dissipation. However, shear strain field in the glass beads bed cannot be measured in this experimental setup.
Clarifying this mechanism remains an important issue for future studies.  

As shown in Fig.~\ref{fig:discuss_image}, a bump is formed in front of the sphere, and we will focus on the relationship between this bump and slip ratio. Even though the sinking depth changes little between rolling up and rolling down, the deformation of the granular surface in front of the sphere is larger when it rolls down a granular slope, as shown in Fig.~\ref{fig:discuss_image}. During rolling down motion, the sphere is supposed to accelerate under gravity but the sphere actually decelerates due to the effect of the bump of glass beads accumulating in front of it. The bumpy structure produces a resistive force opposing its motion. In contrast, during rolling up motion, the deceleration due to gravity dominates over the resistance from the granular bed. These observations indicate that the size of bump is larger during rolling down motion than during rolling up motion. Numerical calculations have shown that the larger the slip ratio of a wheel rolling on granular ground, the smaller the bump formation (terrain deformation formed in front of the wheel)~\cite{Qin_2024}. To purely evaluate the effect of $s$ on this phenomenon, data from different materials could be used. However, material density also affects the sinking depth. To examine the effect of bump formation under identical sinking depths, it is more appropriate to compare data obtained using the same material but different $\alpha$ values. Considering these points, the above-discussed tendency between bump formation and slip ratio is consistent with our observations. To quantitatively discuss this bump formation effect, much more precise measurements of granular surface deformation are required. This is one of the most important issues for future study.

\section{Discussion}\label{sec_Discussion}
In our setup, we observe both translational and rotational motions. In the uphill case studied by Fukumoto et al.~\cite{Fukumoto_2024}, two types of friction were introduced: one arising from the drag force associated with translational motion, and the other arising from the torque associated with rotational motion. For the latter, they introduced a rotational friction term defined from the normalized torque, $I\dot{\omega}$, normalized by $MgR\cos\alpha$, where $I$ is the moment of inertia of a homogeneous sphere with mass $M$ and radius $R$, and $\alpha$ is the slope angle. They found that this rotational friction is almost independent of both the sphere density and the slope angle.

Thus, the present study mainly focuses on the friction coefficient characterizing the translational motion. Specifically, the friction acting on the translational motion is defined through the force balance on the sphere, $Ma_\mathrm{X}$, in which the gravitational component is explicitly taken into account. The equation of motion is given by
\begin{equation}
Ma_\mathrm{X} = -Mg\sin\alpha - F_\mathrm{d},
\label{eq:drag}
\end{equation}
where $a_\mathrm{X}$ is the acceleration of the sphere, $g$ is the gravitational acceleration, and $F_\mathrm{d}$ denotes the drag force excluding gravity. Figure~\ref{fig:vt} shows that the translational motion exhibits nearly constant deceleration. Thus, $F_\mathrm{d}$ can be assumed to be constant. We further assume that $F_\mathrm{d}$ is proportional to the normal force acting on the sphere, $Mg\cos\alpha$. Namely, we introduce $\mu_\mathrm{d}$ as the friction coefficient, which satisfies
\begin{equation}
F_{\mathrm{d}} =\mu_\mathrm{d} Mg\cos\alpha.
\label{eq:mud}
\end{equation} By integrating both sides of the Eq.~(\ref{eq:drag}) and incorporating a correction to the gravitational potential due to the sinking depth $\delta$ and using the relation of Eq.~(\ref{eq:mud}), we obtain the following energy conservation form, 
\begin{equation}
\frac{1}{2}Mv_{0}^2 + Mg[L(-\sin\alpha) + \delta] = \mu_\mathrm{d} MgL\cos\alpha.
\label{eq:energy_conservation}
\end{equation}
This equation implies that the kinetic energy and gravitational potential energy at $X=0$ are assumed to be dissipated in proportion to the travel distance $L$. $\mu_\mathrm{d}$ is computed by substituting the experimental parameters ($\alpha, v_0$) and the measured parameters ($\delta, L$) into Eq.~(\ref{eq:energy_conservation}).

\begin{figure}
    \centering
    \includegraphics[width=.95\columnwidth]{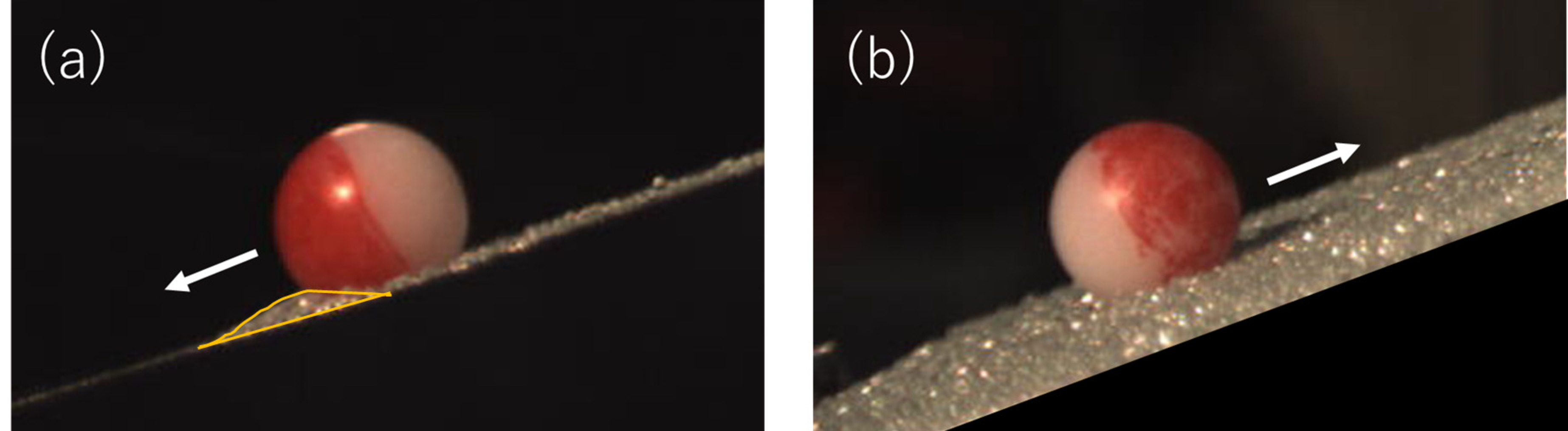}
    \caption{
    The image of the sphere (a) rolling down at $\alpha\simeq-20^{\circ}$, and (b) rolling up at $\alpha\simeq20^{\circ}$ from Fukumoto et al~\cite{Fukumoto_2024}. The radius of polyethylene sphere is 6.35~mm in both (a) and (b). The yellow curve indicates the edge of the bump. The slip ratio $s$ is smaller in (a) and larger in (b).
    }
    \label{fig:discuss_image}
\end{figure}

\begin{figure}
    \centering
    \includegraphics[width=\columnwidth]{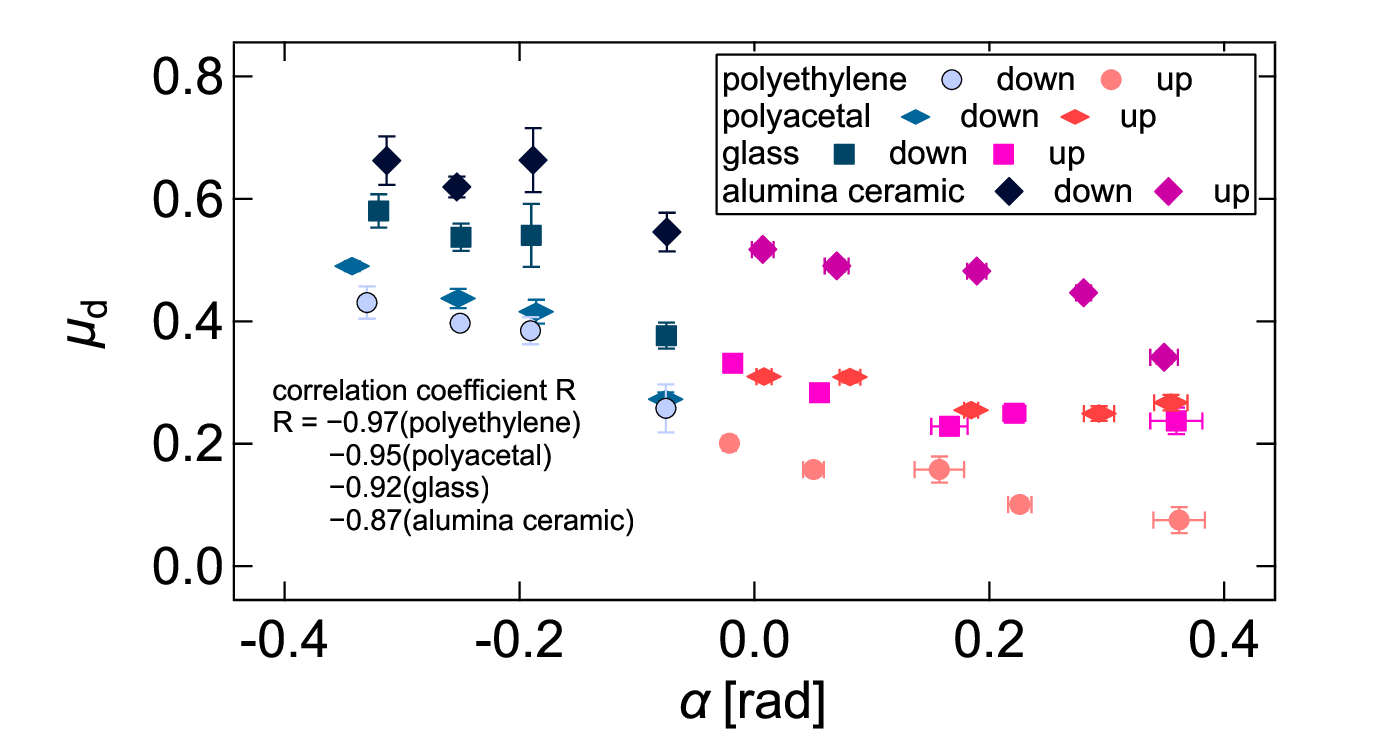}
    \caption{The relation between $\mu_\mathrm{d}$ and angle $\alpha$. $\alpha<0$ and $\alpha>0$ correspond to rolling down and up cases respectively. We use the data ($\alpha>0$) obtained by Fukumoto et al.~\cite{Fukumoto_2024}. Error bars indicate the standard deviation of all the data. R=$-$0.97(polyethylene), $-$0.95(polyacetal), $-$0.92(glass), $-$0.87(alumina ceramic), R: the correlation coefficient.}
    \label{fig:mud_alpha}
\end{figure}

\begin{figure}
    \centering
    \includegraphics[width=.45\columnwidth]{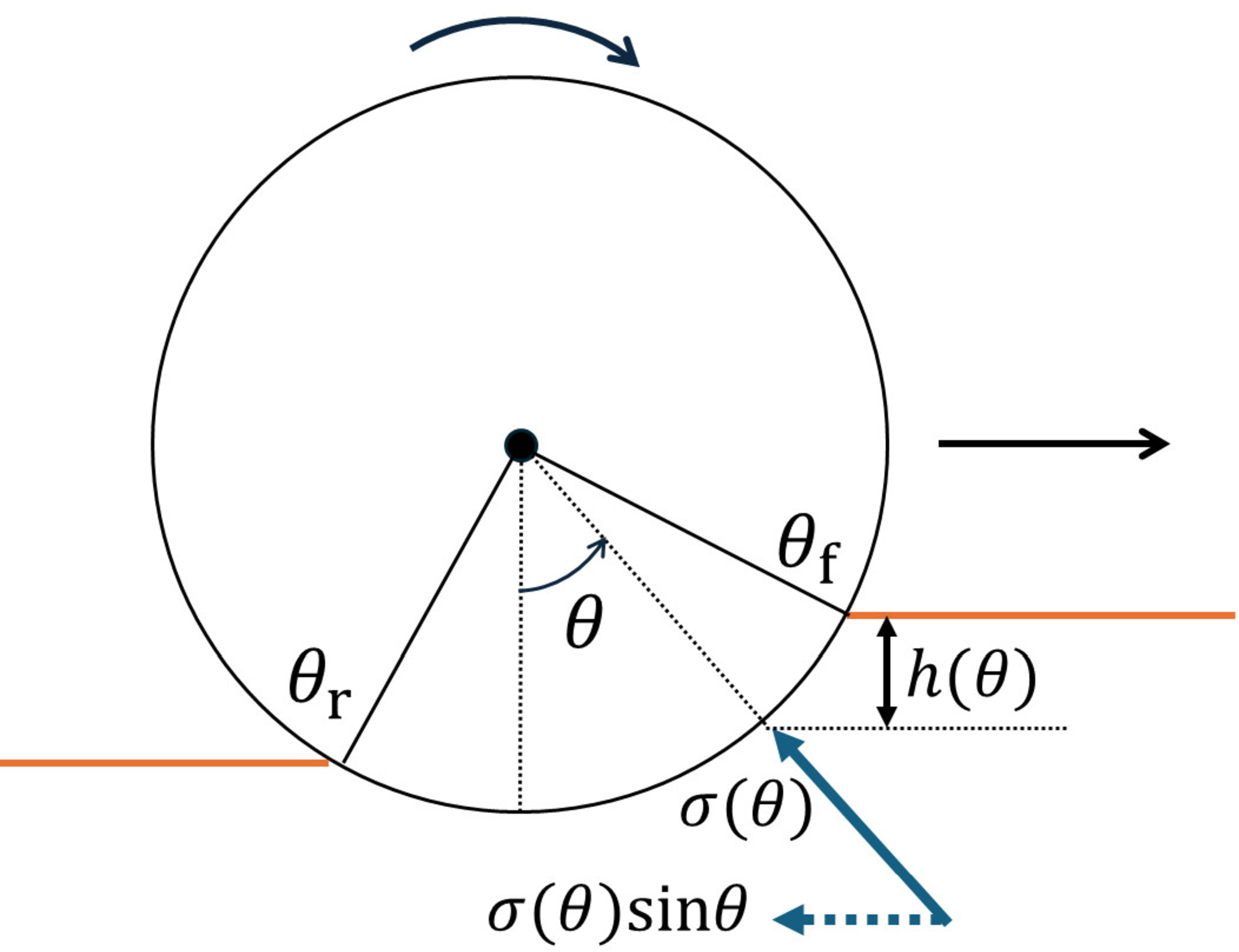}
    \caption{The image of the relation between sinking depth $h(\theta)$ and normal stress $\sigma(\theta)$ when the wheel is rolling on a granular soil. $\theta_\mathrm{f}$ and $\theta_\mathrm{r}$ are the angles of contact with the ground.}
    \label{fig:tera}
\end{figure}

Fig.~\ref{fig:mud_alpha} shows that $\mu_\mathrm{d}$ generally decreases with increasing $\alpha$ as an overall trend. From the correlation degree characterized by the correlation coefficient,  $\mu_\mathrm{d}$ has a clear negative correlation with $\alpha$. This tendency might be qualitatively explained as follows. Figure~\ref{fig:discuss_image} shows that the size of bump is larger during rolling down motion than during rolling up motion, resulting in a larger contact area i.e., larger $\mu_\mathrm{d}$. The relationship between contact area and resistance is qualitatively consistent with the Bekker–Wong–Reece model~\cite{Terramechanics} in terramechanics. In terramechanics models~\cite{Terramechanics}, the resistance force exerted on a wheel with radius $R$ and width $b$ is given by $Rb\int_{\theta_\mathrm{r}}^{\theta_\mathrm{f}}\sigma(\theta)\sin\theta\,d\theta$, where $\theta_\mathrm{f}$ and $\theta_\mathrm{r}$ are the entry and exit angles, respectively, and $\sigma(\theta)$ is the normal stress, as shown in Fig.~\ref{fig:tera}. The relationship between $\sigma(\theta)$ and the sinking depth $h(\theta)$ is empirically known to follow $\sigma(\theta)\propto h(\theta)^n$ with $n>0$. This empirical relation indicates that, as $\theta_\mathrm{f}$ increases due to a larger contact area, the resistance force also increases. Accordingly, in the present study, $\mu_\mathrm{d}$ is expected to increase with increasing contact area due to bump formation at negative slope angle $\alpha$. While the previous study~\cite{Blasio_2009} focused on the decelerating motion of a rolling sphere over a narrow range of slope angles, this study reveals the dependence of $\mu_\mathrm{d}$ on $\alpha$ by examining a wide range of $\alpha$ values in both the uphill~\cite{Fukumoto_2024} and downhill directions.

We investigate the influence of the slip ratio $s$ and the sinking depth $\delta$ on the friction coefficient $\mu_\mathrm{d}$, which directly affects the motion of the sphere because $\alpha$ is a geometric parameter. In the following, we directly examine the relationship between $\mu_\mathrm{d}$ and $s$ through their respective dependencies on $\alpha$, as demonstrated below.

\begin{figure}
    \centering
    \includegraphics[width=\columnwidth]{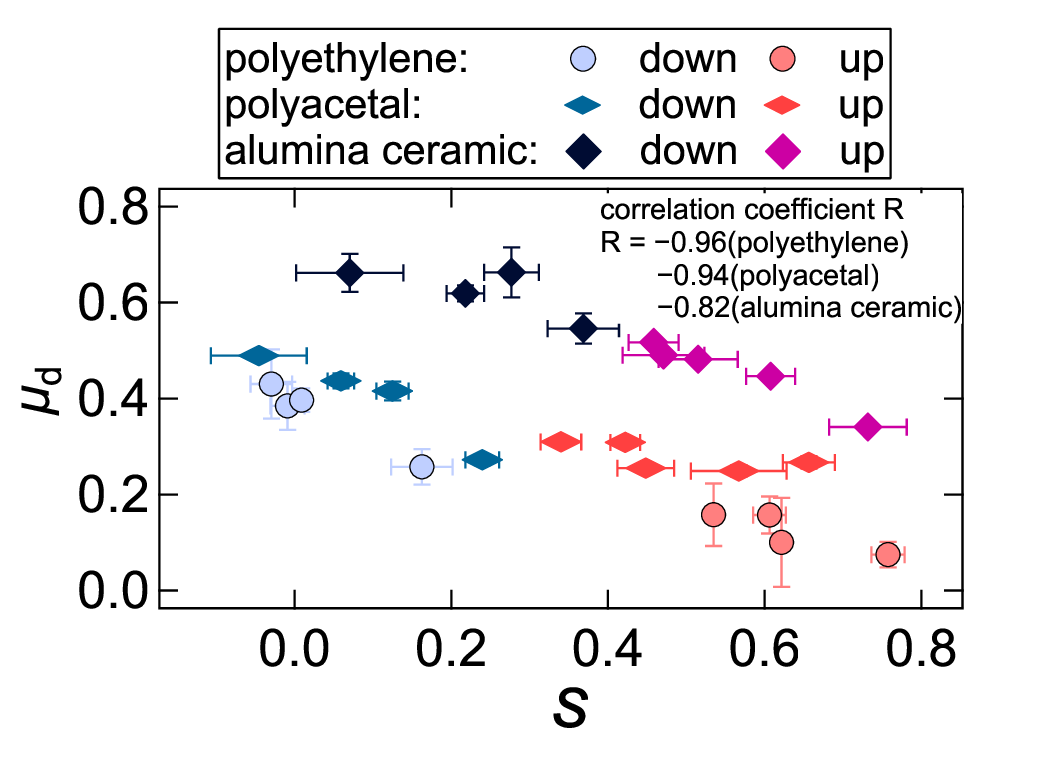}
    \caption{The relation between $\mu_\mathrm{d}$ and slip ratio $s$, except for the glass sphere, at each $\alpha$. Error bars indicate the standard deviation of all the data. The data of rolling up is taken from Fukumoto et al.~\cite{Fukumoto_2024}. R=$-$0.96(polyethylene), $-$0.94(polyacetal), $-$0.82(alumina ceramic), R: the correlation coefficient.}
    \label{fig:mud_slip}
\end{figure}

As shown in Fig.~\ref{fig:mud_slip}, $\mu_\mathrm{d}$ roughly decreases linearly with $s$ as an overall trend. The correlation coefficient values suggest the clear negative correlation between $\mu_\mathrm{d}$ and $s$.
This result suggests that, in the regime of high slip ratio $s$, rotational motion is partly converted into a driving force for translational motion. Seguin conducted numerical simulations of the drag force exerted on a wedge-shaped intruder moving in a granular medium~\cite{Seguin_2022}. Their results showed that the drag force decreases and rotational torque increases with increasing rotational speed, indicating a transfer of mechanical power from translational motion to rotational motion. A similar mechanism may also be present in the present experiment.

\begin{figure}
    \centering
    \includegraphics[width=\columnwidth]{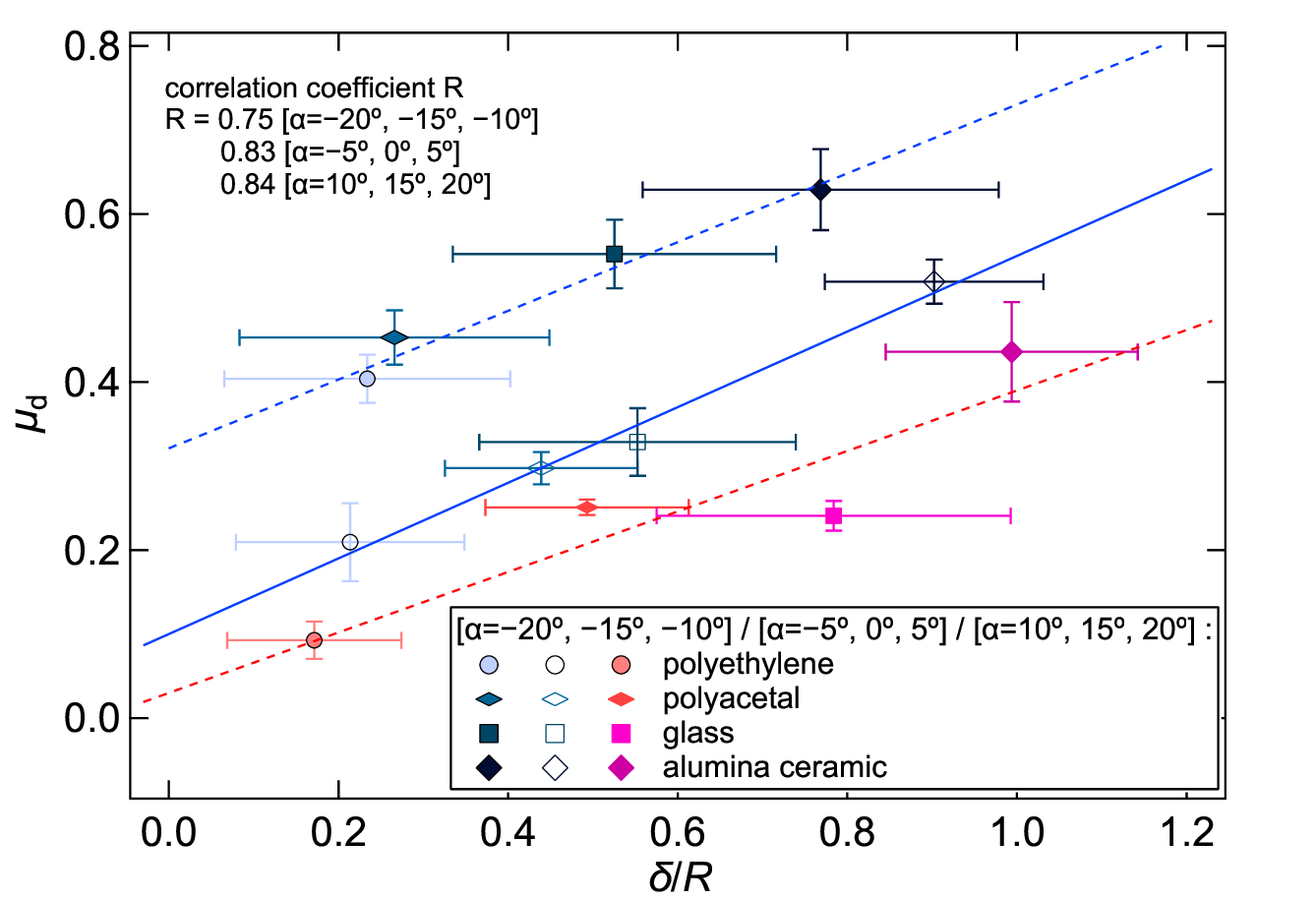}
    \caption{
    The relation between $\mu_\mathrm{d}$ and $\delta/R$. Error bars indicate the standard deviation of all the data. Filled blue, empty blue, and filled red symbols represent the data rolling down the granular slop for [$\alpha=-20^{\circ}, -15^{\circ}, -10^{\circ}$], [$\alpha=-5^{\circ},  0^{\circ}, 5^{\circ}$], [$\alpha=10^{\circ}, 15^{\circ}, 20^{\circ}$], respectively. Blue solid and dashed lines indicate the least-square linear fitting line [Eq.~(\ref{eq:mud_line})] with the slope of $\beta=0.41\pm0.04$ and $0.45\pm0.04$ with the intercept of $\mu_0=0.32\pm0.02$ and $0.1\pm0.02$ respectively. The red dashed line indicates the least-square fitting line with the slope of $\beta=0.36\pm0.1$ with the intercept of $\mu_0=0.03\pm0.07$. The fitting error is the standard deviation of the least-square fitting. We use the data ($\alpha>0$) obtained from~\cite{Fukumoto_2024}. R=0.75 [$\alpha=-20^{\circ}, -15^{\circ}, -10^{\circ}$], 0.83[$\alpha=-5^{\circ},  0^{\circ}, 5^{\circ}$], 0.84[$\alpha=10^{\circ}, 15^{\circ}, 20^{\circ}$], R: the correlation coefficient.
    }
    \label{fig:mud_d}
\end{figure}

As shown in Fig.~\ref{fig:fitting_depth}, $\delta$ increases with $\rho_\mathrm{s}$. In addition, Fig.~\ref{fig:mud_alpha} shows that $\mu_\mathrm{d}$ also increases roughly with $\rho_\mathrm{s}$. Therefore, we investigate the relationship between $\mu_\mathrm{d}$ and $\delta/R$. It is difficult to obtain a clear relation between $\mu_\mathrm{d}$ and $\delta/R$ for each individual angle $\alpha$ because of the limitations in measurement accuracy. To address this, we divide $\alpha$ into three groups that exhibit similar values of $\mu_\mathrm{d}$: [$\alpha=-20^{\circ}, -15^{\circ}, -10^{\circ}$], [$\alpha=-5^{\circ}, 0^{\circ}, 5^{\circ}$], and [$\alpha=10^{\circ}, 15^{\circ}, 20^{\circ}$]. For each group, we compute the average values of $\mu_\mathrm{d}$ and $\delta/R$ and analyze the relationship between them, as shown in Fig.~\ref{fig:mud_d}. The correlation coefficient values show that $\mu_\mathrm{d}$ has a clear positive correlation with $\delta/R$.

Figure~\ref{fig:mud_d} shows that $\mu_\mathrm{d}$ increases approximately linearly with $\delta/R$. Thus, we assume that $\mu_\mathrm{d}$ obeys
\begin{equation}
\mu_\mathrm{d} = \beta(\delta/R) + \mu_0,
\label{eq:mud_line}
\end{equation}
where $\beta$ is the proportionality coefficient and $\mu_0$ is the intercept. This linear trend is also consistent with the uphill case~\cite{Fukumoto_2024}. When the least-square fitting is applied to the $\mu_\mathrm{d}$ and $\delta/R$ for each $\alpha$ group, the slope of the fitted line does not show a statistically significant difference ($\beta \simeq 0.41$). In contrast, the intercept $\mu_0$ is statistically larger in the downhill group [$\alpha=-20^{\circ}, -15^{\circ}, -10^{\circ}$] than in the other groups.

$\mu_\mathrm{d}$ can be regarded as a friction coefficient that characterizes the energy dissipation mainly caused by the deformation of the granular bed. A larger deformation $\delta$ leads to a larger contact area between the sphere and the granular bed, and therefore $\mu_\mathrm{d}$ increases. Consequently, a positive correlation is observed between $\mu_\mathrm{d}$ and $\delta$. Moreover, the intercept $\mu_0$ increases as $\alpha$ decreases. As shown in Fig.~\ref{fig:mud_slip}, the slip ratio $s$ also decreases as $\alpha$ decreases, suggesting a negative correlation between the intercept $\mu_0$ and $s$. In the following, we investigate the relationship between $\mu_0$ and $s$.

\begin{figure}
    \centering
    \includegraphics[width=\columnwidth]{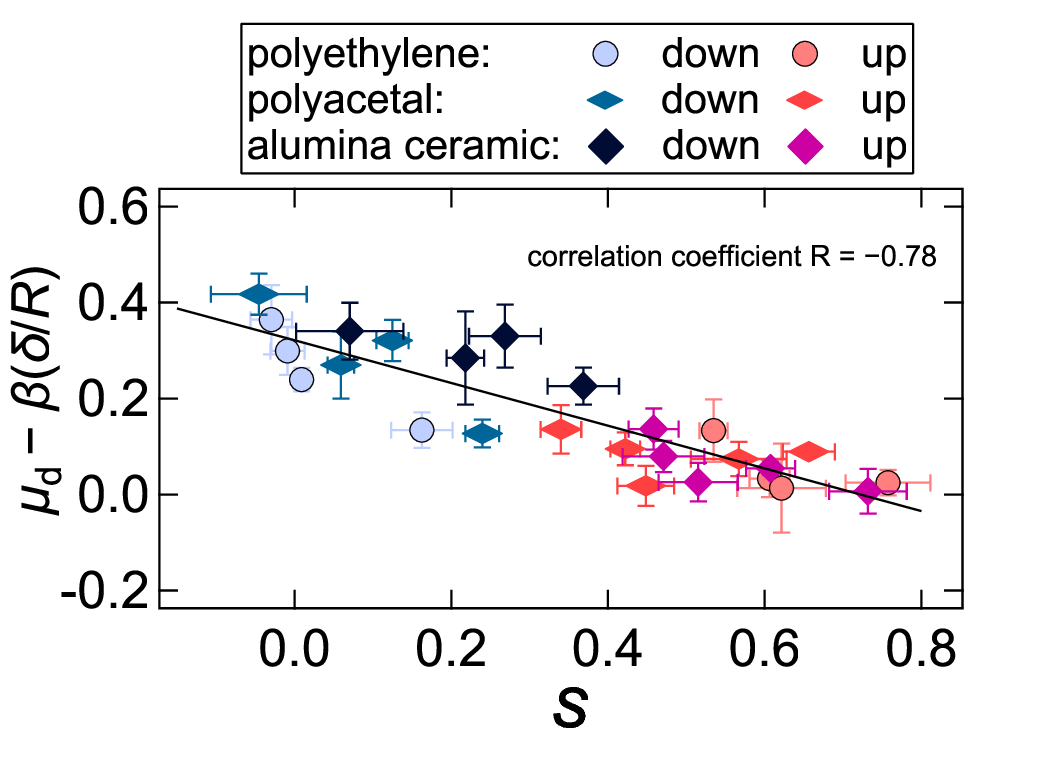}
    \caption{The relation between $\mu_\mathrm{d} - \beta(\delta/R)$ and $s$ at each $\alpha$ ($\beta=0.41$). The black solid line indicates the least-square linear fitting line, $\mu_\mathrm{d} - \beta(\delta/R)=0.32-0.44s$. Error bars indicate the standard deviation of all the data. The correlation coefficient R, R=$-$0.78.
    }
    \label{fig:mu_ds}
\end{figure}

Specifically, we investigate the relationship between $\mu_0 ,(= \mu_\mathrm{d} - \beta(\delta/R))$ and $s$, as shown in Fig.~\ref{fig:mu_ds}. The correlation coefficient values show that $\mu_0$ has a clear negative correlation with $s$. The value of $\mu_0$ decreases linearly with $s$, as indicated in Fig.~\ref{fig:mu_ds}. By applying least-square fitting to the data in Fig.~\ref{fig:mu_ds}, we obtain the empirical relation, $\mu_\mathrm{d}-\beta(\delta/R) = 0.32-0.44s~(\beta=0.41)$. To the best of our knowledge, this study provides the first quantitative empirical formula linking $\mu_\mathrm{d}$, $s$, and $\delta/R$. This relation suggests that $\mu_\mathrm{d}$ consists of two contributions: one associated with the sinking deformation and the other associated with the bump formation in front of the sphere, which is related to the slip ratio $s$. The frictional contribution due to sinking is determined solely by the density of the sphere and is independent of the angle $\alpha$. In contrast, the contribution associated with bump formation correlates with the slip ratio $s$, which depends on the angle $\alpha$.

We compare the current results with previous studies~\cite{Blasio_2009,Stefaan_2017,Texier_2018}. In the downhill study conducted by Blasio and Saeter~\cite{Blasio_2009}, $\delta/R\simeq0.3$ and $\mu_\mathrm{d}\simeq0.45$, which fit this empirical formula [Eq.~(\ref{eq:mud_line})] if the slip ratio was similar to that in our study. This suggests that the empirical formula for downhill rolling may also be applicable to other conditions. In the rolling experiments on a flat granular surface conducted by Wal et al.~\cite{Stefaan_2017}, $\delta/R\simeq0.01$ and $0.03$, and the corresponding values of $\mu_\mathrm{d}$ were $0.0524$ and $0.0655$, respectively, under no-slip conditions. In the downhill experiments conducted under no-slip conditions by Texier et al.~\cite{Texier_2018}, $\delta/R\simeq0.05$ and $\mu_\mathrm{d}\simeq0.2$. These results suggest that the empirical formula [Eq.~(\ref{eq:mud_line})] does not apply under conditions of shallow subsidence depths. Determining the range of applicability of the empirical formula obtained here will be an important task for future work. 

In this study, our objective is to investigate the interaction between a rolling object and deformable granular media by simplifying the system and eliminating complex phenomena. To achieve this, we use dry grains in order to exclude the effects of adhesion and grain aggregation. In addition, we employed spheres with nearly identical surface conditions (i.e., similar static friction coefficients), so that sphere density would serve as the primary variable rather than surface properties. We focus on macroscopic interactions as a first step toward understanding the mechanics between a sphere and granular media. A more detailed investigation at the microscopic level would further advance understanding of the underlying interaction mechanisms. Although the present study intentionally eliminates complex factors, it would be interesting in future work to examine the effect of surface conditions by using actively adhering spheres, or to investigate how changes in substrate properties such as particle density and size distribution affect the dynamics.

In this research, it is assumed that the rigidity of the spheres and granular materials does not affect the behavior of the spheres, but the case where the rigidity of each material has an effect is also an interesting issue that should be solved in the future.

In this study, we conduct downhill rolling experiments across a wide range of sphere densities and slope angles. By comparing the result with uphill results~\cite{Fukumoto_2024}, we find that $\mu_\mathrm{d}$ depends on $\alpha$, $s$, and $\delta$. However, since in the experimental system, $s$ and $\delta$ are not controlled independently, it is not straightforward to isolate their individual influences on $\mu_\mathrm{d}$. Developing methods to investigate $\mu_\mathrm{d}$ using $s$ and $\delta$ as input parameters is an important issue to be addressed in future work.

\section{Conclusion}\label{sec_conclusion}
In previous studies, no universal law was established for the relationship between effective friction and sinking depth that applies to both rolling-up and rolling-down cases. In this study, we characterize the friction of a sphere rolling down a granular slope. The dynamics are experimentally investigated by varying the initial velocity, slope angle, and sphere density. We selected a parameter range that was not explored in previous studies of rolling-down motion. The sinking depth scales as, $\delta/R = C_{\rho}\,(\rho_\mathrm{s}/\rho_\mathrm{g})^{3/4},$ which is consistent with previous results for uphill rolling~\cite{Fukumoto_2024}. The translational motion exhibits constant deceleration. To characterize the motion, we introduce the effective friction coefficient $\mu_\mathrm{d}$. $\mu_\mathrm{d}$ shows a decreasing tendency with the slope angle $\alpha$ and the slip ratio $s$. Also, $\mu_\mathrm{d}$ increases linearly $\delta/R$, $\mu_\mathrm{d}=\beta(\delta/R)+\mu_0$. $\beta\simeq0.41$ does not vary significantly, while the value of $\mu_0$ is linearly decreasing with $s$. Understanding why $\mu_\mathrm{d}$ takes larger values in the downhill direction, while maintaining almost the same proportional slope $\beta$, remains a crucial issue for future work.

%
%


\ack{This work was supported by JST SPRING, Grant No. JPMJSP2138, JSPS-DST Bilateral Program, Grant No. JPJSBP120227710, and JST ERATO, Grant No. JPMJER2401.}


\data{Further information
is available in the online guidelines: https://doi.org/10.5281/zenodo.18810364}

\suppdata{The typical examples of actual movies}


\bibliographystyle{iopart-num} 
\bibliography{take}

@ARTICLE{Sanderson_2010,
   author       = "K. Sanderson",
   title        = "Mars rover spirit(2003-2010)",
   year         = "2010",
   journal      = "Nature",
   volume       = "463",
   pages        = "600",
}

@ARTICLE{escape_2019,
  author={Zhao, Xuan and Liu, Pan and Yu, Qiang and Shi, Peilong and Ye, Yiming},
  journal={IEEE Access}, 
  title={On the Effective Speed Control Characteristics of a Truck Escape Ramp Based on the Discrete Element Method}, 
  year={2019},
  volume={7},
  number={},
  pages={80366-80379},
}

@article{dungbeetle, 
title={Rules for the Leg Coordination of Dung Beetle Ball Rolling Behaviour}, 
volume={10}, 

journal={Scientific Reports}, 
author={Leung, Binggwong. and Bijma, Nienke and Baird, Emily and Dacke, Marie and Gorb, Stanislav and Manoonpong, Poramate}, 
year={2020}, 
pages={9278} 
}

@article{pill_bug,
    author = {Smigel, Jacob T. and Gibbs, Allen G.},
    title = {Conglobation in the pill bug, Armadillidium vulgare, as a water conservation mechanism},
    journal = {Journal of Insect Science},
    volume = {8},
    number = {1},
    pages = {44},
    year = {2008},
}

@article{Texier_2018,
year = {2018},
month = {oct},
publisher = {EDP Sciences, IOP Publishing and Società Italiana di Fisica},
volume = {123},
pages = {54005},
author = {Darbois Texier, B. and Ibarra, A. and Vivanco, F. and Bico, J. and Melo, F.},
title = {Friction of a sphere rolling down a granular slope},
journal = {Europhysics Letters},
}

@article{Terramechanics,
title = {Prediction of rigid wheel performance based on the analysis of soil-wheel stresses part I. Performance of driven rigid wheels},
journal = {Journal of Terramechanics},
volume = {4},
pages = {81},
year = {1967},
author = {Jo-Yung Wong and A.R. Reece}
}

@article{Uehara_2003,
  title = {Low-Speed Impact Craters in Loose Granular Media},
  author = {Uehara, J. S. and Ambroso, M. A. and Ojha, R. P. and Durian, D. J.},
  journal = {Phys. Rev. Lett.},
  volume = {90},
  
  pages = {194301},
  numpages = {4},
  year = {2003},
  month = {May},
  publisher = {American Physical Society}
}

@article{Fukumoto_2024,
  title = {Energy dissipation of a sphere rolling up a granular slope: Slip and deformation of the granular surface},
  author = {Fukumoto, T. and Yamamoto, K. and Katsura, M. and Katsuragi, H.},
  journal = {Phys. Rev. E},
  volume = {109},
  
  pages = {014903},
  numpages = {8},
  year = {2024},
  month = {Jan},
  publisher = {American Physical Society}
}

@article{Rinse_2020,
    author = {Liefferink, Rinse W. and Aliasgari, Mojgan and Maleki-Jirsaraei, Nahid and Rouhani, Shahin and Bonn, Daniel},
    title = {Sliding on wet sand},
    journal = {Granular Matter},
    year = {2020},
    volume = {22},
    pages = {57}
}

@article{Liefferink_2018,
  title = {Ploughing friction on wet and dry sand},
  author = {Liefferink, R. W. and Weber, B. and Bonn, D.},
  journal = {Phys. Rev. E},
  volume = {98},
 
  pages = {052903},
  numpages = {5},
  year = {2018},
  month = {Nov},
  publisher = {American Physical Society}
}

@article{Carvalho_2024,
  title = {Drag reduction during the side-by-side motion of a pair of intruders in a granular medium},
  author = {Carvalho, D. D. and Bertho, Y. and Seguin, A. and Franklin, E. M. and Texier, B. Darbois},
  journal = {Phys. Rev. Fluids},
  volume = {9},
  
  pages = {114303},
  numpages = {13},
  year = {2024},
  month = {Nov},
  publisher = {American Physical Society}
}

@article{Crassous_2017,
  title = {Pressure-Dependent Friction on Granular Slopes Close to Avalanche},
  author = {Crassous, J\'er\^ome and Humeau, Antoine and Boury, Samuel and Casas, J\'er\^ome},
  journal = {Phys. Rev. Lett.},
  volume = {119},
  
  pages = {058003},
  numpages = {5},
  year = {2017},
  month = {Aug},
  publisher = {American Physical Society}
}

@article{Katsuragi_2007,
    author = {Katsuragi, Hiroaki and Durian, Douglas J.},
    title = {Unified force law for granular impact cratering},
    journal = {Nature Physics},
    year = {2007},
    volume = {3},
    pages = {420}
}

@article{Jung_2017,
    author = {Jung, Wonjong and Choi, Sung Mok and Kim, Wonjung and Kim, Ho-Young},
    title = {Reduction of granular drag inspired by self-burrowing rotary seeds},
    journal = {Physics of Fluids},
    volume = {29},
    pages = {041702},
    year = {2017},
    month = {04},
}

@article{Seguin_2022,
  title = {Forces on an intruder combining translation and rotation in granular media},
  author = {Seguin, A.},
  journal = {Phys. Rev. Fluids},
  volume = {7},
  
  pages = {034302},
  numpages = {11},
  year = {2022},
  month = {Mar},
  publisher = {American Physical Society}
}

@article{Seguin_2013,
  title = {Experimental velocity fields and forces for a cylinder penetrating into a granular medium},
  author = {Seguin, A. and Bertho, Y. and Martinez, F. and Crassous, J. and Gondret, P.},
  journal = {Phys. Rev. E},
  volume = {87},

  pages = {012201},
  numpages = {10},
  year = {2013},
  month = {Jan},
  publisher = {American Physical Society}
}

@article{Ding_2011,
  title = {Drag Induced Lift in Granular Media},
  author = {Ding, Yang and Gravish, Nick and Goldman, Daniel I.},
  journal = {Phys. Rev. Lett.},
  volume = {106},
 
  pages = {028001},
  numpages = {4},
  year = {2011},
  month = {Jan},
  publisher = {American Physical Society}
}

@article{Darbois_2017,
  title = {Helical Locomotion in a Granular Medium},
  author = {Darbois Texier, Baptiste and Ibarra, Alejandro and Melo, Francisco},
  journal = {Phys. Rev. Lett.},
  volume = {119},
  
  pages = {068003},
  numpages = {5},
  year = {2017},
  month = {Aug},
  publisher = {American Physical Society}
}

@article{Gondret_2022,
  title = {Added-mass force in dry granular matter},
  author = {Seguin, A. and Gondret, P.},
  journal = {Phys. Rev. E},
  volume = {105},
  
  pages = {054903},
  numpages = {5},
  year = {2022},
  month = {May},
  publisher = {American Physical Society}
}

@article{Brzinski_2013,
  title = {Depth-Dependent Resistance of Granular Media to Vertical Penetration},
  author = {Brzinski, T. A. and Mayor, P. and Durian, D. J.},
  journal = {Phys. Rev. Lett.},
  volume = {111},

  pages = {168002},
  numpages = {5},
  year = {2013},
  month = {Oct},
  publisher = {American Physical Society}
}

@article{Pacheco_2009,
    author = {Pacheco-V\'azquez, F. and Ruiz-Su\'arez, J. C.} ,
    title = {Sliding through a superlight granular medium},
    journal = {Phys.\ Rev.\ E},
    year = {2009},
    volume = {80},
    pages    = {060301(R))},
}

@ARTICLE{Stefaan_2017,
   author       = "Stefaan Van Wal and Simon Tardivel and Paul Sánchez and Darius Djafari-Rouhani and Daniel Scheeres ",
   title        = "Rolling resistance of a sphrical pod on a granular bed",
   year         = "2017",
   journal      = "Granular Matter",
   volume       = "19",
   pages        = "17",
}

@ARTICLE{Blasio_2009,
   author       = "De Blasio, Fabio Vittorio and Saeter, May-Britt",
   title        = "Rolling friction on a granular medium",
   year         = "2009",
   journal      = "Phys.\ Rev.\ E",
   volume       = "79",
   pages        = "022301",
}

@article{Ikeda_2022,
author = {Ikeda, Ayame and Kumagai, Hiroyuki and Morota, Tomokatsu},
title = {Topographic Degradation Processes of Lunar Crater Walls Inferred From Boulder Falls},
journal = {Journal of Geophysical Research: Planets},
volume = {127},
pages = {e2021JE007176},
year = {2022}
}

@article{Bickel_2021,
author = {Bickel, V. T. and Aaron, J. and Manconi, A. and Loew, S.},
title = {Global Drivers and Transport Mechanisms of Lunar Rockfalls},
journal = {Journal of Geophysical Research: Planets},
volume = {126},
number = {10},
pages = {e2021JE006824},
year = {2021},
}

@article{Cui_2017,
    author = {Cui, Sheng-hua and Pei, Xiang-jun and Huang, Run-qiu},
    title = {Rolling motion behavior of rockfall on gentle slope: an experimental approach},
    journal = {Journal of Mountain Science} ,
    year = {2017},
    volume = {14},
    pages = {8}
}

@article{Seguin_2019,
    author ={Seguin, A.} ,
    title = {Hysteresis of the drag force of an intruder moving into a granular medium},
    journal = {The European Physical Journal E},
    year = {2019},
    volume = {42},
    pages = {13}
}

@article{Potiguar_2013,
  title = {Lift and drag in intruders moving through hydrostatic granular media at high speeds},
  author = {Potiguar, Fabricio Q. and Ding, Yang},
  journal = {Phys. Rev. E},
  volume = {88},
  pages = {012204},
  numpages = {11},
  year = {2013},
  month = {Jul},
  publisher = {American Physical Society}
}

@article{Okubo_2022,
	author = {Okubo, Fumiaki and Katsuragi, Hiroaki},
	title = {Impact drag force exerted on a projectile penetrating into a hierarchical granular bed},
	journal = {A\&A},
	year = {2022},
	volume = {664},
	pages = "A147",
}

@article{iikawa,
    author = {Iikawa, Naoki and Katsuragi, Hiroaki},
    journal = {Acta Geotechnica},
    year = {2025},
    title = {Resistive force modeling for shallow cone penetration into dry and wet granular layers},
    pages = {1279},
    volume = {20}
}

@article{Qin_2024,
title = {DEM simulation and continuation algorithm of granular physical field for planetary wheel-terrain interaction},
journal = {Powder Technology},
volume = {433},
pages = {119197},
year = {2024},
author = {Qingning Lan and Zhengyin Wang and Liang Ding and Huaiguang Yang and Haibo Gao and Lutz Richter and Zongquan Deng},
}

@article{shear,
  title = {Strain localization in dry sheared granular materials: A compactivity-based approach},
  author = {Ma, Xiao and Elbanna, Ahmed},
  journal = {Phys. Rev. E},
  volume = {98},
  pages = {022906},
  numpages = {18},
  year = {2018},
}

\end{document}